\documentclass[usenatbib,onecolumn]{mn2e}

\title[Weak-field Schwarzschild geodesics]
{An analytic solution for weak-field Schwarzschild geodesics}

\author[Daniel J. D'Orazio and Prasenjit Saha]
{Daniel J. D'Orazio\thanks{dorazio@physik.uzh.ch} and
Prasenjit Saha\thanks{psaha@physik.uzh.ch }\\
Institute for Theoretical Physics, University of Z\"urich, 8057 Z\"urich,
Switzerland}

\begin{document}

 \date{Accepted 2010 April 20. Received 2010 April 16; in original form 2010 March 26}

\pagerange{\pageref{firstpage}--\pageref{lastpage}} \pubyear{2010}

\maketitle

\label{firstpage}

\begin{abstract}
  It is well known that the classical gravitational two body problem
  can be transformed into a spherical harmonic oscillator by
  regularization. We find that a modification of the regularization
  transformation has a similar result to leading order in general
  relativity. In the resulting harmonic oscillator, the leading-order
  relativistic perturbation is formally a negative centrifugal
  force. The net centrifugal force changes sign at three Schwarzschild
  radii, which interestingly mimics the innermost stable circular
  orbit (ISCO) of the full Schwarzschild problem. Transforming the
  harmonic-oscillator solution back to spatial coordinates yields, for
  both timelike and null weak-field Schwarzschild geodesics, a
  solution for $t,r,\phi$ in terms of elementary functions of a
  variable that can be interpreted as a generalized eccentric anomaly.
  The textbook expressions for relativistic precession and light
  deflection are easily recovered. We suggest how this solution could
  be combined with additional perturbations into numerical methods
  suitable for applications such as relativistic accretion or dynamics
  of the Galactic-centre stars.
\end{abstract}


\section{Introduction}
When Schwarzschild geodesics appear in classic tests of general
relativity, the important result is an integral over the geodesic:
orbital precession or deflection of light.  Similary, in modern tests
of relativity involving binary pulsars \citep[for a review,
see][]{lrr-2006-3} the observable effects are also cumulative over
many orbits.

In the case of the recently-discovered S-stars near the Galactic
Centre, the cumulative effects of relativity are no longer the
principal quantity of interest. The highly eccentric examples from the
S-stars \citep{2008ApJ...689.1044G, 2009ApJ...692.1075G}, which experience a
range of gravitational regimes, motivate an interest in tracking
relativistic effects as they vary along an orbit.  In particular, some
recent work has drawn attention to relativistic effects on redshifts
near pericentre passage
\citep{2006ApJ...639L..21Z,2009ApJ...690.1553K,2010ApJ...711..157A}.
These effects can be calculated numerically, and some of them
also by post-Newtonian perturbation theory, but a simpler method is
desirable.

Such a method is suggested by Levi-Civita or Kustaanheimo-Stiefel (LC
or KS) regularization, which are transformations of the classical
gravitational two-body problem to an equivalent harmonic oscillator.
This type of regularization was originally introduced in two
dimensions \citep{1920LC} and much later extended to three dimensions
\citep{1964Kustaanheimo,1965KS}.  KS regularization has an extensive
literature, including applications to $N$-body simulations
\citep{1974CeMec..10..185A,1974CeMec..10..516A,1989ApJS...71..871J,1993CeMDA..57..439M}.
The classical result suggests that the LC or KS regularization could
be used to transform the general relativistic problem into a perturbed
harmonic-oscillator.  We find even better: a modification of the LC/KS
transformation acting on the geodesics  of the leading order Schwarzschild
metric in the isotropic or harmonic gauge \citep[cf.][\unskip~Section~8.2]{1972gcpa.book.....W}, 
\begin{equation}
  ds^2 = - \left(
           1 - {2M\over r} + {2M^2\over r^2} + O\left(M^3\over r^3\right)
           \right) dt^2
         + \left(
            1 + {2M\over r} + O\left(M^2\over r^2\right)
           \right) d{\bf x}^2,
\label{metric}
\end{equation}
yields an unperturbed circular/spherical harmonic oscillator.  As a result, the solution is
analytic.  The difference from the classical case is a negative
centrifugal-force term in the transformed space. This terms encodes
the leading-order effects of precession, deflection of light, and the
innermost stable orbit.

Because orbits in the Schwarzschild spacetime do not leave the orbital
plane, in this paper we mainly consider the two-dimensional or LC
case.  The three-dimensional or KS case is similar, but algebraically
more complicated, as it involves introducing a fourth spatial
dimension.

\section{Transformation of the Geodesic Equations}
Since regularization is formulated in the language of Hamiltonians, we
begin by expressing the geodesic equations in Hamiltonian form.  A
convenient expression for the Hamiltonian is \citep[e.g., Equation
25.10 of][]{1973grav.book.....M}
\begin{equation}
  H = {1\over{2}}g^{\mu \nu} p_{\mu} p_{\nu}.   
\end{equation}  
Considering the metric (\ref{metric}), we have
\begin{equation}
  H = -{1\over{2}}\left(1 + {{2M}\over{r}} +
    {{{2M^2}}\over{r^{2}}}\right) {p_t}^{2} + 
  \left(1- {{2M}\over{r}} \right){p^2 \over{2}},
  \label{hamilt} 
\end{equation}
This retains the leading-order Newtonian corrections, which are
$O(M/r)$ in the spatial part and $O(M^2/r^2)$ in the temporal part,
and neglects higher orders. Note that the Hamiltonian (\ref{hamilt})
has four degrees of freedom: time $t$ being a coordinate and $p_t$ its
conjugate momentum, while the affine parameter $\lambda$ is the
independent variable.  We assume units with $G/c^2=1$.

The Hamiltonian being independent of $t$, it follows that $p_t$ is a
constant, and Hamilton's equation for $t$ reads
\begin{equation} 
{dH\over{dp_t}} = {dt\over{d\lambda}} = -\left(1 +
    {2M\over{r}} + {{2M^2}\over{r^2}} \right) p_t.
\label{timeq}
\end{equation}
where $\lambda$ denotes the affine parameter.  On choosing $p_t = -1$
(which we are free to do, as this amounts to choosing units for
$\lambda$) and discarding a constant we arrive at the Hamiltonian
\begin{equation}
  H = -{1\over{2}}\left( {2M\over{r}} + {{2M^2}\over{r^2}} \right)
      + \left(1-{2M\over{r}} \right){p^2 \over{2}}.
\label{hamilt2}
\end{equation} 
As is well known from the spherical symmetry of the Schwarzschild
spacetime, geodesics are confined to a plane. Without loss of
generality we can choose a two dimensional coordinate system in the
$x,y$ plane. (Below, at the end of this section, we briefly indicate
the procedure without this simplification.)

We now apply a regularization transformation.  In the $x,y$ plane, we
introduce two new coordinates, given by the real and imaginary parts
of the complex number
\begin{equation}
  Q = \sqrt{x+iy}.
\end{equation}
Hence $r=|Q|^2$.  The conjugate momentum components are given by the
real and imaginary parts of a complex number $P$, which satisfies
\begin{equation}
  p_x + ip_y = {{Q^*}{P} \over{2|Q|^2}}.
\end{equation}
On multiplying the above equation by its complex conjugate we arrive
at the transformation
\begin{equation}
  p^2 = {|P|^2 \over 4|Q|^2}.
\label{ptrans}
\end{equation}
The $P,Q$ are known as Levi-Civita (or LC) variables. A proof that
they are indeed canonical appears in several sources \citep[for
example][]{2009MNRAS.400..228S} and we do not repeat it here.
Transforming (\ref{hamilt2}) and rearranging yield the Hamiltonian in
the LC variables
\begin{equation}
  H = -{1\over{2}}\left( {2M\over{|Q|^2}} + {{2M^2}\over{|Q|^4}} \right) +
  \left({1\over{|Q|^2}} - {2M\over{|Q|^4}} \right){|P|^2\over{8}}.
\end{equation} 
To complete the regularization we invoke a Poincar\'e time
transformation which involves introducing a scaled time variable
related to the affine parameter by
\begin{equation}
  d\lambda=g(P,Q)\,ds,
\label{timetrans}
\end{equation}
where $g(P,Q)$ can be any function, and define a new Hamiltonian
\begin{equation}
  \Gamma \equiv g\,\left( H - E \right).
\label{Geq}
\end{equation}
The $\Gamma$ Hamiltonian (\ref{Geq}) preserves the Hamiltonian form of
the equations of motion, provided the constant $E$ is the initial
value of the original Hamiltonian.  We choose
\begin{equation}
  g = \left({1\over{|Q|^2}} - {{2M}\over{|Q|^4}}\right)^{-1} = {
    |Q|^2\over{1 - 2M/|Q|^2} },
  \label{g} 
\end{equation}
which approximates to $|Q|^2 + 2M$ for large $|Q|$.  The
time-transformed Hamiltonian takes the form
\begin{equation}
  \Gamma = {|P|^2\over{8}} - E|Q|^2 - {3}{{M^2}\over{|Q|^2}} -
  M(1+2E) + O \left( |Q|^{-4} \right).
  \label{hamilt3} 
\end{equation}

We remark that the classical case uses $g=|Q|^2$, and only the first
two terms in (\ref{hamilt3}) are present in the $\Gamma$ Hamiltonian,
up to a constant.  However, as is evident from (\ref{hamilt3}), the
weak-field Schwarzschild case also yields a harmonic oscillator under
a regularization transformation. The essential difference is in the
extra $1/|Q|^2$ term which encodes the leading order effects of
general relativity by altering the classical angular momentum in
(\ref{hamilt3}).  Thus, the key modification from the classical case
which allows the formulation of an analytically solvable, relativistic
Hamiltonian is the choice of $g(P,Q)$ in equation (\ref{g}).

If we do not restrict the coordinates to the plane, the LC
transformation must be replaced with a Kustaanheimo-Steifel
transformation. Although the transformation itself is far more
complicated (involving four spatial dimensions) the time
transformation and $\Gamma$ Hamiltonian (\ref{hamilt3}) remain the
same, except that $|P|$ and $|Q|$ are lengths in four Euclidean
dimensions.  This is easily seen on comparing the above with KS
regularization of the Kepler problem \citep[see, e.g., Section~5
of][]{2009MNRAS.400..228S}.

\section{Solutions}
Continuing in two dimensions, it is possible to put the $\Gamma$
Hamiltonian in a more recognizable form by transforming to polar
coordinates $(Q_r, Q_{\phi}, P_r, P_{\phi})$,
\begin{equation}
  \Gamma = {{P^2_r}\over{8}}
  + {\left( P^2_{\phi} - 6{(2M)^2} \right)\over{8Q^2_r}}
  - E{Q^2_r} - M(1+2E) + O\left( Q^{-4}_r \right),
\label{hamiltpolar}
\end{equation}
which for negative values of $E$ is the Hamiltonian for a circular classical
harmonic oscillator with squared angular momentum decreased by
$6(2M)^2$ from the equivalent Kepler problem.

We remark that the $\Gamma$ Hamiltonian appears to depend only on
terms up to $1/r$, and it would seem that by leaving out terms of
order $|Q|^{-4}$ we have omitted relativistic effects. However, this
is not the case, since the time equation now hides a factor of
$r$. Thus, one should read terms of order $|Q|^{-n}$ as terms of order
$r^{-(n/2+1)}$ for $n=0,1,2\ldots$ This provides some insight as to why
the Hamiltonian now has soluble equations of motion: we have pushed a
factor of $r$ into the time equation.

We introduce the constants
\begin{equation}
  M' \equiv M(1+2E) \qquad \mbox{and}  \qquad
  {P'}^2_{\phi} \equiv P^2_{\phi} - 6(2M)^2.
\end{equation}
Dropping higher order terms, the Hamiltonian becomes,
\begin{equation}
  \Gamma = {{P^2_r}\over{8}} + {{P'}^2_{\phi} \over{8Q^2_r}}
  - E{Q^2_r} - M' ,
\end{equation}
which is identical to the transformed Hamiltonian of a particle with
angular momentum $P'_{\phi}$ in a central force potential of the form
$V(r) = M'/r$, or equivalently to the Kepler problem where $M$ is
replaced by $M'$ and $P^2_{\phi}$ is replaced by ${P'}^2_{\phi}$.

Hamilton's equations of motion are
\begin{eqnarray}
 \label{Heqs} 
  {dQ_r\over{ds}} &=& {P_r\over{4}} \nonumber \\
  {dP_r\over{ds}} &=&  2E{Q_r} + {{P'}^2_{\phi} \over{4}}
  {1\over{{Q^3_r}}} \\
  {dQ_{\phi}\over{ds}} &=& {P_{\phi}\over{4}}
  {1\over{{Q^2_r}}} \nonumber
\end{eqnarray}
while $P_{\phi}$ is a constant.  On combining with the time
transformation (\ref{timetrans}), the equation (\ref{timeq}) for the
time coordinate becomes
\begin{eqnarray}
  dt &=& \left( Q^2_r + {2M}
    +{{2M^2}\over{Q^2_r}} \right) {ds\over{1-2M/Q^2_r}} \nonumber \\
  &\simeq& \left[4M + {Q^2_r} + 6{M^2\over{Q^2_r}} + O \left(
      {{Q^{-4}_r}} \right) \right]ds ,
 \label{teqn} 
\end{eqnarray}
where we have again used the approximation $Q^2_r \gg 2M$.

We remark that ${P}_{\phi}=2p_{\phi}$, which is the weak-field
relationship between LC angular momentum and angular momentum in
standard coordinates. This can be seen from a comparison of the third
of equations (\ref{Heqs}) and the corresponding Hamilton equation
applied to (\ref{hamilt2}) in the time variable $s$.

\subsection{Bound orbits}
For negative values of $E$ the solutions are bound orbits. We define a
`classical' and a `relativistic' semi-major axis
\begin{equation}
  a\equiv{M\over{2|E|}} \qquad   a'\equiv{M'\over{2|E|}}, 
\end{equation}
the eccentric anomaly,
\begin{equation}
  \beta = \sqrt{2|E|}s
\end{equation} 
and the eccentricity
\begin{equation}
  e=\sqrt{ 1 + {{P'}^2_{\phi}E \over{2 M'^2}} }
\end{equation} 
in the standard way. Solving equation (\ref{teqn}) gives an implicit
equation for the coordinate time $t$ in terms of $\beta$ and solving
equations (\ref{Heqs}) via quadrature gives the following solutions
for the LC variables in terms of elementary functions of $\beta$.
\begin{eqnarray}    	\label{solnb}
  {Q^2_{r}}(\beta) &=& a' \left[ 1 - e \cos{ \beta } \right]
  \nonumber \\
  {Q_{\phi}}(\beta) &=& {P_{\phi}\over{P'_{\phi}}} \mbox{tan}^{-1}
  \left[ \sqrt{{ 1+e \over{ 1-e }}} \tan{\left( { \beta \over{2} }
      \right)} \right] \\
  t_{} &=& \sqrt{{{a'}^3\over{M'}}} \left[ \beta - e\sin{\beta} \right] +
  {4M\over{\sqrt{2|E|}}} \beta + {6(2M)^2\over{P_{\phi}}}Q_{\phi}, \nonumber
\end{eqnarray}
where we have chosen $Q_{\phi}(0) = 0$, and $Q^2_r(0) = a' (1-e) =
r_{\rm min}$.  The LC radial momentum $P_r(\beta)$ is then generated
from the first of equations (\ref{Heqs}) and, with the above choice
of $Q^2_r(0)$, the initial radial momentum vanishes.

Note that in terms of the classical semi-major axis of the Kepler
problem the quantity $a' = a - M$. So it is simple to show that
equations (\ref{solnb}) reduce to the Kepler LC equations of motion
when $M$ is small.

\subsection{Unbound trajectories}
For the unbound case the Hamiltonian and thus the equations of motion
only change by the sign of $E$ which appears in $a'$, $e$, and
$\beta$. With this substitution the unbound solutions become
\begin{eqnarray}           \label{solnu} 
  Q^2_r(\beta) &=& a' \left[ e\cosh{ \beta } - 1 \right] 
  \nonumber \\
  Q_\phi(\beta) &=& {P_{\phi}\over{P'_{\phi}}} \tan^{-1} 
  \left[  \sqrt{{ e+1  \over{ e-1 }}}  \mbox{tanh}\left( { \beta \over{2} }
    \right) \right]  \\
  t &=& \sqrt{{{a'}^3\over{M'}}} \left[ e\sinh{\beta} - \beta \right]
  +  {4M\over{\sqrt{2|E|}}} \beta +
  {6(2M)^2\over{P_{\phi}}}Q_{\phi} , \nonumber
\end{eqnarray}
where we have chosen $Q^2_r(0) = a' \left( e-1 \right)$, which in
the unbound case is the point of closest approach.

\subsection{Light rays}
Null geodesics are the solutions to Hamilton's equations when $H$ in
equation (\ref{hamilt}) is set to zero. Since we discarded a constant
of $-1/2$ in the derivation of the LC Hamiltonian, the null solutions
can be found by assigning $E$ in the unbound case the value of
$1/2$. This amounts to redefining the constants $a'$, $\beta$, and $e$
in equations (\ref{solnu}) such that the null equations of motion
become
\begin{eqnarray}   	\label{solnull} 
  Q^2_r(s) &=& 2M \left[ e_n\cosh{ s } - 1 \right] 
  \nonumber \\
  Q_\phi(s) &=& {P_{\phi}\over{P'_{\phi}}} \tan^{-1} 
  \left[  \sqrt{{ e_n+1  \over{ e_n-1 }}}  \mbox{tanh}\left( { s \over{2} }
    \right) \right]  \\
  t &=& {2M} \left[ e_n\sinh{s} - s \right]
  +  4M  s + {6(2M)^2\over{P_{\phi}}}Q_\phi , \nonumber
\end{eqnarray}
with
\begin{equation}
  e_n = \sqrt{ 1 + {{P'}^2_{\phi}\over{16M^2}} }  \label{enull} 
\end{equation}
and $Q^2_r(0) = 2M (e_n - 1)$.

\section{Properties of the solution}
Now that we have derived the bound, unbound, and null equtions of
motion we show that, to first order, they reproduce the predictions for
geodesics in a Schwarzschild spacetime, namely those of orbital
precession, the deflection of light, and the innermost stable
circular orbit (ISCO).

\subsection{Orbital precession}
The prefactor of the middle line in (\ref{solnb}) automatically gives
the precession rate of orbits and tells us that the precession is due
to the $-6(2M)^2$ perturbation of the classical squared angular
momentum. This is equivalent to the conventional interpretation
of precession being caused by an additional centrifugal force
term. Substituting for $P'_{\phi}$ we have
\begin{equation} 
  {P_{\phi}\over{P'_{\phi}}} = \left[ 1 - 6
    {(2M)^2\over{P^2_{\phi}}} \right]^{-1/2}.
\end{equation}
Since $ P^2_{\phi} \propto Q^4_r$, then $P^2_{\phi} \gg (2M)^2$ and we may write
\begin{equation}
  \left[ 1 - 6 {(2M)^2\over{P^2_{\phi}}} \right]^{-1/2} \simeq 1 + 3{(2M)^2\over{P^2_{\phi}}}.
\end{equation} 
We note that, due to the complex square root nature of the LC
transformation, $Q_{\phi} = {1\over{2}}\phi $. Thus as the solution
evolves through one period $Q_{\phi}$ will increase by
\begin{equation}
  \pi + 3\pi{(2M)^2\over{P^2_{\phi}}} \mbox{  } \mbox{rad}
\end{equation} 
giving an orbital precession rate of
\begin{equation}
  \triangle Q_{\phi} = 3\pi{(2M)^2\over{P^2_{\phi}}} \mbox{  } {\mbox{rad}\over{\mbox{orbit}}}.
\end{equation}
Converting back to non-LC coordinates and expressing the previous
equation in terms of the semi-latus rectum of the orbit $\alpha =
a\left( 1 - e \right)$, where $a$ is the classical semi-major axis,
the precession rate becomes
\begin{equation}
  \triangle\phi = 6\pi{MG\over{c^2\alpha}} \mbox{  } {\mbox{rad.}\over{\mbox{orbit}}},
\end{equation}
as in the conventional treatment \citep[e.g.,][]{1972gcpa.book.....W}.

\subsection{Deflection of light}
We may similarly derive the deflection angle of a light ray. Imagine
that the light ray starts infinitely far from the Schwarzschild mass
at $t=-\infty$, approaches the point of closest approach at $t=0$, and
continues on to $t=\infty$. If there is no deflection $Q_{\phi}$ will
sweep out $\pi/2$ radians since $Q_{\phi} = {1\over{2}}\phi$. If there
is deflection the total difference in $Q_{\phi}(-\infty)$ and
$Q_{\phi}(\infty)$ will be greater than ${\pi\over{2}}$. To
determine the deflection angle we compute,
\begin{equation}
  \triangle Q_\phi =  2\left[ Q_\phi(\infty) -
    Q_\phi(0)\right]  - {\pi\over{2}} = 2{P_{\phi}\over{P'_{\phi}}}
  \tan^{-1}\left[ {e_n+1\over{\sqrt{e^2_n-1}}} \right] - 0 - {\pi\over{2}},
\end{equation}
where $Q_\phi$  is the null solution for $Q_{\phi}$. Substituting for
$P'_{\phi}$ and the expression for $e_n$ in equation
(\ref{enull}), invoking the binomial approximation for ${P'}^2_{\phi}
\gg M^2$ and $P^2_{\phi} \gg M^2$, and neglecting terms of order $M^2$
and greater we find
\begin{equation}
  \triangle Q_\phi =  2 \left(1 + O\left(M^2\right) \right)
  \tan^{-1}\left[ 1 + {4M\over{P'_{\phi}}} + O\left( M^2
    \right) \right] + 0 - {\pi\over{2}}.
\end{equation}
Using the expansion 
\begin{equation}
  \tan^{-1}\left( 1 + x \right) = {\pi\over{4}} + {1\over{2}}x +  O\left( x^2 \right) 
\end{equation}
we obtain 
\begin{equation}
  \triangle Q_\phi = {\pi\over{2}} + 2\left[ {2M\over{P'_{\phi}}} -
    O(M^2) \right] - {\pi\over{2}}. 
\end{equation}
Converting back to non-LC coordinates we gain a factor of $1/2$ since
the LC angular momentum $P'_{\phi}$ is twice the non-LC angular
momentum $p'_{\phi}$. We may treat $p'_{\phi}$ as the corrected
(relativistic) angular momentum which is equal to the impact parameter
$b$ for photons (with the speed of light set to unity).  Thus we
arrive at the expression, to first order in $M$, for the angle by
which light is deflected due to a spherically symmetric mass
distribution,
\begin{equation}
  \triangle \phi_{\rm light} = {4MG\over{b}} \simeq {4MG\over{r_0}}. 
\end{equation}
Where to this order the impact parameter can be replaced by $r_0$, the
point of closest approach.
\subsection{Innermost stable orbits}
Due to the weak-field approximation used to derive the geodesic
equations of motion, the solutions do not exhibit an event
horizon. Remarkably though, the solutions do reproduce the phenomenon
of an innermost stable circular orbit.  This is due to the centrifugal
term ${6(2M)^2/{Q^2_r}}$ in the LC Hamiltonian (\ref{hamiltpolar}).

From the first two of Hamilton's equations (\ref{Heqs}) for a bound
orbit we find
\begin{equation} 
  {d^2Q_r\over{ds^2}} = - {|E|\over{2}}{Q_r} \left[1 -
    {{P'}^2_{\phi} \over{8|E|Q^4_r}}\right].
\label{ISCO1}
\end{equation} 
and notice that equation (\ref{ISCO1}) becomes a one dimensional
harmonic oscillator in the radial coordinate (i.e., the condition for
radial free fall, or equivalently $e\rightarrow 1$) when
${P'}^2_{\phi}=0$. Writing out $P'_{\phi}$, the condition becomes
\begin{equation}
  P^2_{\phi} = 6(2M)^2.
\end{equation}
Dividing by $|E|$ and using the definitions of the classical semi-major
($a$) and semi-minor ($b$) axes of the orbit this condition becomes
\begin{equation}
  b^2 = 6aM.
\end{equation}
Which in the case of a circular orbit ($a=b$) is precisely the ISCO predicted
by the conventional analysis of timelike geodesics in a Schwarzschild
metric. Note that although this relation has been derived for the
classical semi-major axis one can convert to the standard radial
coordiante of the Schwarzschild metric and find an identical result as
follows. First convert to the relativistic semi-major axis,
$a'=a-M$ which, for the above ISCO, gives $a'=5M$. Then recall that the
standard Schwarzschild radial coordinate $R$ is given in terms of
the isotropic radial coordinate by (see e.g. Weinberg 1972 section
8.2)
\begin{equation}
  R=r\left( 1+{M\over{2r}}\right)^2 \simeq r\left( 1+{M\over{r}}\right),
\end{equation}
where the approximation is made to be consistent with the derivation
of the equations of motion. For $a'=r=5M$ this gives $R=6M$ as
expected.

To interpret the ISCO derived here we observe from equation
(\ref{ISCO1}) that the condition for zero radial acceleration is
\begin{equation} 
  Q^2_r = \sqrt{P^2_{\phi} - 6(2M)^2 \over{8|E|}}.
\label{ISCO2}
\end{equation} 
Now consider a classical orbit with a fixed $ P^2_{\phi}$ and imagine
`turning on' relativity adiabatically keeping $P^2_{\phi}$
constant. As relativity is turned on the negative centrifugal force
term increases and effectively reduces ${P'}^2_{\phi}$. If $P^2_{\phi}
> 6(2M)^2$ to begin with, then the classical orbit will shrink by the
appropriate amount in the presence of relativistic effects. But if we
originally had $P^2_{\phi} \leq 6(2M)^2$, then as relativity is turned
on equation (\ref{ISCO2}) shows that $Q^2_r$ would shrink to 0
implying that all orbits for which $P^2_{\phi} \leq 6(2M)^2$ are
unstable.

Also note that the standard coordinate singularity at $R=2M$ is mapped
to $r=M$ in the aproximate isoptropic coordinates. This yields the
following interpretation for the the LC equations of motion.
Substituting $a'=a-M$ in the first of equations (\ref{solnb}) for the LC radial
coordinate and restricting to the circular case where $e=0$ we find
\begin{equation} 
  {Q^2_r}(\beta) + M = a,
\end{equation}
which has the interpretation that the LC radial coordinate is measured
not from the origin of the corresponding Kepler problem but from the
event horizon predicted by the Schwarzschild metric. This does not
carry the interpretation of a horizon since when $e\neq0$ orbits can
still come arbitrarily close to $r=0$ and continue outside of
$r=M$. However, it does add an interesting interpretation to the
regularization transformation. For circular orbits the regularization essentially cuts out the area
inside the event horizon and stitches it back together mapping a
circle to the origin. For elliptical orbits one can imagine an
analogous interpretation.

Since these equations of motion predict the correct ISCO and exhibit
special behavior at the Schwarzschild event horizon, perhaps they
could be particularly useful for simple approximate modelling of
relativistic accretion discs.

\subsection{Error terms}
Working backwards from the approximate Hamiltonian in the time
variable $s$ (equation \ref{hamilt3}) by solving equation (\ref{Geq})
for $H$, we can recover the error terms in our approximate Hamiltonian.
Solving
\begin{equation}
  \Gamma = {|Q|^2\over{1-2M/|Q|^2}} (H - E)
  \label{gherr}
\end{equation}
with no further approximations and converting to non-LC coordinates
yields what we may call a surrogate Hamiltonian
\begin{equation}
  H_{\rm surr} = -{1\over{2}}\left[ 1+{{2M}\over{r}} +
    {2M^2\over{r^{2}}} - 
    \left( {12M^3\over{r^3}} \right)  \right] {p_t}^{2} +
  \left(1- {{2M}\over{r}} \right){p^{2} \over{2}} +
  {4M^2\over{r^2}}E. 
  \label{hamilterr}
\end{equation}
for which the analytic solution is exact.  For timelike geodesics $E$
is small and for null geodesics $1/r^2$ is a second order
correction. Thus to obtain a first order expression for error in the
metric components we may neglect the term containing $E$.

If it is bothersome that the Hamiltonian in equation (\ref{hamilterr})
is dependent on $E$, one can eliminate it by a further modification.
Let us add a $E/{|Q|^4}$ term to the $\Gamma$ Hamiltonian,
\begin{equation}
\Gamma' = \Gamma +  {4M^2\over{|Q|^4}}E,
\end{equation}
which makes only a higher-order change to the solutions.
Introducing $H'_{\rm surr}$ by
\begin{equation}
  \Gamma' = |Q|^2 (1 + 2M/|Q|^2 + 4M^2/|Q|^4) \, (H'_{\rm surr}-E)
\end{equation}
one finds
\begin{equation}
H'_{\rm surr} = -{1\over{2}}
         \left[ { 1 + 4M/r + 10M^2/r^2 \over 
                  1 + 2M/r + 4M^2/r^2  } \right]
         {p_t}^{2}
         + \left[  { 1 \over { 1 + 2M/r + 4M^2/r^2 } } \right]
         {p^{2}\over{2}}
\label{hamilterr2}
\end{equation}
which, when expanded in $1/r$ to the appropriate order, is identical
to (\ref{hamilterr}) less the term dependent on $E$. The Hamiltonian
(\ref{hamilterr2}) is then the Hamiltonian for which equations
(\ref{solnb}), (\ref{solnu}), and (\ref{solnull}) are exact solutions.


\section{Discussion}

We have derived timelike and null geodesics in the leading-order
Schwarzschild metric in terms of elementary functions.  The expressions
(\ref{solnb}) for bound orbits and (\ref{solnu}) for unbound orbits,
together with (\ref{solnull}) for light rays, are all simple
generalizations of well-known expressions in classical celestial
mechanics.  The usual formulas for relativistic orbital precession and
light deflection are easily recovered.  A feature resembling the
innermost stable circular orbit in the full Schwarzschild metric is
also present.

The technique we have used is a modification of the Levi-Civita or
Kustaanheimo-Stiefel regularization transformation and transforms the
geodesic equation into a spherical harmonic oscillator.  The
simplicity of the result, notwithstanding the non-trivial route used
to derive it, hints at some underlying symmetry in the Schwarzschild
problem.  We speculate that it is somehow related to the separability
of the Hamilton-Jacobi and other equations in the Schwarzschild and
Kerr metrics \citep[cf.][]{1983mtbh.book.....C} but have not attempted
to investigate this.

As mentioned in the Introduction, the original motivation for this
work was to find useful formulas applicable to the highly-eccentric
Galactic-centre stars, whose orbits pass through a large range of
gravitational regimes.  Future observations of these stars aiming to detect
relativistic effects will require computation of relativistic effects
on both stellar orbits and light rays at many points along an orbit, 
for many orbits, in order to fit the orbital parameters.  The solutions in this paper allow a simpler, more efficient method for carrying out those computations. The
analytic solutions will not be sufficient on their own because the
Galactic-centre stars also experience additional Newtonian
perturbations due to local matter \citep{2008AJ....135.2398M}, but
they can be incorporated into numerical methods, specifically,
generalized leapfrog integrators.  Such algorithms evolve alternately
under two Hamiltonians, which are integrable separately.  The idea
goes back to \cite{1991AJ....102.1528W} and
\cite{1991CeMDA..50...59K}. Some recent developments on adaptive
stepsizes appear in \cite{2007CeMDA..98..191E} and are applied to the
specific problem of Galactic-centre stars in
\cite{2009ApJ...703.1743P}.  We note, however, that the present work
is limited to test particles, and hence will not be applicable
for binary orbits or self-gravitating disc simulations unless a
generalization is found.

Another potential application may be the use of the solutions in relativistic disc
simulations as an alternative to the widely used pseudo-Newtonian
potentials \citep[see
especially][]{1980A&A....88...23P,1996ApJ...461..565A,2009A&A...500..213A},
an advantage being that the solutions in this paper are well-defined approximations and
include a more complete repertoire of general-relativistic effects for
the same computational budget.

\section{Acknowledgements}
We thank Raymond Ang\'elil for discussion and comments. We also thank Swagata Nandi and the referee for pointing out errors in an earlier version of the manuscript.  DJD was supported by a Fulbright-Swiss Government Fellowship.

\def\apj{ApJ}
\def\apjl{ApJ}
\def\apjs{ApJS}
\def\aap{A\&A}
\def\mnras{MNRAS}
\def\aj{AJ}

\bibliographystyle{mn2e}
\bibliography{ms.bbl}

\bsp

\label{lastpage}
\end{document}